# Human Mobility Trends during the COVID-19 Pandemic in the United States

Minha Lee[1], Jun Zhao[1], Qianqian Sun[1], Yixuan Pan[1], Weiyi Zhou[1], Chenfeng Xiong[1], Lei Zhang[1*]

[1] Maryland Transportation Institute, Department of Civil and Environmental Engineering,
University of Maryland
1173 Glenn Martin Hall, College Park, MD 20742 USA

[*] Corresponding author:
Dr. Lei Zhang, Herbert Rabin Distinguished Professor
Director, Maryland Transportation Institute
Email: lei@umd.edu

**Abstract**
In March of this year, COVID-19 was declared a pandemic and it continues to threaten public health. This global health crisis imposes limitations on daily movements, which have deteriorated every sector in our society. Understanding public reactions to the virus and the non-pharmaceutical interventions should be of great help to fight COVID-19 in a strategic way. We aim to provide tangible evidence of the human mobility trends by comparing the day-by-day variations across the U.S. Large-scale public mobility at an aggregated level is observed by leveraging mobile device location data and the measures related to social distancing. Our study captures spatial and temporal heterogeneity as well as the sociodemographic variations regarding the pandemic propagation and the non-pharmaceutical interventions. All mobility metrics adapted capture decreased public movements after the national emergency declaration. The population staying home has increased in all states and becomes more stable after the stay-at-home order with a smaller range of fluctuation. There exists overall mobility heterogeneity between the income or population density groups. The public had been taking active responses, voluntarily staying home more, to the in-state confirmed cases while the stay-at-home orders stabilize the variations. The study suggests that the public mobility trends conform with the government message urging to stay home. We anticipate our data-driven analysis offers integrated perspectives and serves as evidence to raise public awareness and, consequently, reinforce the importance of social distancing while assisting policymakers.

# Introduction

Historically, the first half of 2020 will be remembered for the global battle against an invisible enemy. Since the emergence of the novel coronavirus (COVID-19) in December 2019 in Wuhan, China, the world is experiencing unprecedented phenomena.[1] In March of this year, COVID-19 was declared a pandemic by the World Health Organization (WHO), and emergency measures have been internationally implemented as the outbreak continues to threaten public health. As of April 10, 2020, there were almost 1·7 million worldwide confirmed cases of COVID-19, with the United States accounting for over 500,000 cases, or around 30% of overall infections around the world.[2] As a result, over 40 American states have instituted stay-at-home orders, making quarantine and social distancing the new norm for the majority of the U.S. population.

Interdisciplinary research has been actively conducted to mitigate the spread of COVID-19 and its adverse impacts on society. Epidemiologic measurements have been explored to identify the dynamics of disease regarding the spread risk and the effect of human mobility.[3–10] In addition, since the mobility restrictions are considered as a critical factor to prevent the disease spread, studies have assessed its impact.[11–16] Non-pharmaceutical observations are also proven to be effective data sources to have an integrated perspective. Especially big data allow for increased understanding of human behavior changes in response to the spread of the virus. One recent study captures the dissemination of COVID-19 information in relation to the outbreak progression from crowdsourced data.[17] Other studies employing data-driven methodologies have also been introduced to estimate the negative impacts on various sectors such as the economy, public health, and human mobility.[18–22] In particular, mobility data have been identified as being especially relevant and researchers have provided in-depth knowledge on how to leverage mobile device location data for analyzing COVID-19 propagation.[23–25] Technology companies have presented insights on mobility trends by exploiting location data as well.[26–28]

As revealed from literature, there exists the importance of social distancing and timely decision on mobility interventions to slow the pandemic. The key factor to stop the virus is those classical approaches including social distancing, quarantine, and mobility interventions since no treatment is currently available.[13] However, little is known about the gap between social distance advocacy and the actual practices among the general populace.[23] There also has been scant research on public reactions to COVID-19 as well as the interventions on mobility. Realizing the urgent needs on understanding mobility trends amid the pandemic, as one of the pioneering big data-driven studies on COVID-19, we aim to quantify changes in human mobility to provide tangible and intuitive evidence on individual and governmental efforts to migrate the spread. The goal of the study is two-fold. First, we explore large-scale public mobility patterns and the existence of heterogeneity across the nation by leveraging mobile device location data. Our study covers: temporal trend analysis before and after the emergence of COVID-19 and mobility interventions; geospatial trend analysis at national and state levels in the U.S.; and groupwise comparisons regarding sociodemographic characteristics. Second, we uncover potential research areas that can greatly contribute to the current and potential future matters with the observed evidence. While this paper does not intend to establish a complete guidance on how governments or similar bodies should respond, the hope is to share our observations and findings to provide an integrated perspective on the public mobility reactions before and during the pandemic.

# Methods

### Data sources

The University of Maryland COVID-19 Impact Analysis Platform aggregates mobile device location data from more than 100 million anonymized sample devices each month.[29] The aggregated location data are then integrated with COVID-19 case data from John Hopkins University and census population data to monitor the mobility trends in United States. The metrics produced from the data are provided only in aggregated forms at the county, state, and national levels.

### Data analysis

The research team first integrates and preprocesses person and vehicle movement data to improve the quality of our mobile device location data panel, followed by the trip identification process (Figure 1). Second, location points are clustered into activity locations, while home and work locations are identified at census block group (CBG) level. Third, additional trip information including trip origin, destination, departure time, and arrival time are identified based on previously developed and validated algorithms.[30] The condition of staying at home is defined if an anonymized individual in the sample does not travel farther than one mile from home. In the next step, we expand the sample data to the population level by incorporating a multi-level weighting procedure to have results that represent the entire population in a nation, state, or county. The data sources and algorithms implemented are validated based on various datasets such as the National Household Travel Survey (NHTS) and American Community Survey (ACS) and previously peer-reviewed by an external expert panel.[30] Lastly, mobility metrics are integrated with other data sources, such as COVID-19 cases and population. Table 1 summarizes the metrics we adapt in this study while the whole set of metrics can be consulted in the platform.[29] Additional details of the methodology can be found in a separate paper by the authors.[31]

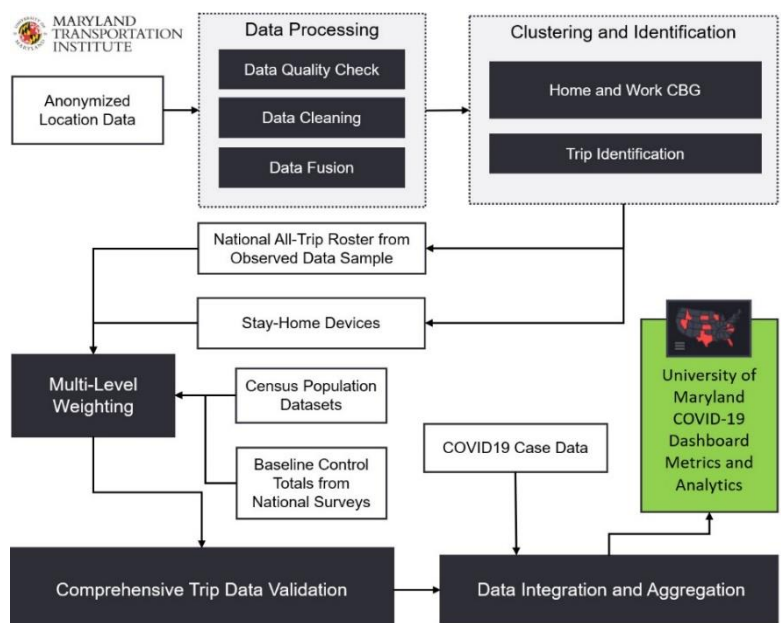

*Figure 1:* **Data Processing and Methodology Framework**[30]

*Table 1:* **A Summary of Metrics Adapted from the COVID-19 Impact Analysis Platform**

| Metrics | Description | Nation | State | Group |
|---|---|---|---|---|
| % staying home | Percentage of residents staying at home | √ | √ | √ |
| miles traveled per person | Average person-miles traveled on all modes | √ | √ | √ |
| % out-of-county trips | The percent of all trips taken that travel out of a county | √ | | √ |
| number of trips per person | Average number of trips taken per person | √ | | |
| number of work trips/person | Number of work trips per person | √ | | √ |
| number of non-work trips/person | Number of non-work trips per person | √ | | |
| number of COVID-19 cases | Number of confirmed COVID-19 cases from the Johns Hopkins University's GitHub repository[32] | √ | √ | √ |

**Spatiotemporal Trend analysis**
We explore the mobility variations regarding the COVID-19 progression and government stay-at-home orders by applying the metrics that are closely related to social distancing. Our trend analysis design can be categorized into three types: 1) nationwide; 2) statewide; and 3) sociodemographic groupwise and the metrics applied are marked in Table 1, respectively. The temporal range covers from January 6, 2020 to April 9, 2020, while weekends are excluded to eliminate noises.

The nationwide trends are examined by applying a 3-day moving average method for all mobility metrics. The statewide trends compare 50 states with two types of timelines. The first is universal timelines: 1) benchmark week (February 3 - February 16) and the most recent week (April 6 - April 12). All states are considered in these timelines. The second type is stay-at-home order timelines: 1) one week before the order and 2) one week after the order, which vary per state and are applied to 44 states with the order implemented as of April 2. Then the statewide trend analysis further evaluates the public reaction stability based on one measure, the percentage of people staying home, which we believe to have a high correlation with social distancing. The stability is measured with a variance, where a higher variance indicates lower stability.

The sociodemographic groupwise comparison endeavors how these features influence the mobility patterns amid COVID-19. During the preliminary analysis, four features are considered: percentage of middle-aged population (35 years old and over); percentage of elderly people (65 years old and over); median household income; and population density in persons per mile of land area. States are classified into two groups by each feature (higher: 25 states and lower: 26 states). In this paper, the result section delivers two comparison results from the median income and population density groups, which show a notable clustering nature, whereas the other two age-related features do not.

**Results**
**Nationwide trends**
A large number of people have decreased their daily movements: the percentage of people staying home rapidly increases from 20% on normal days (benchmark week) to 35% after the outbreak (most recent week); out-of-county trips decreases from 28% to 23%; average trip distance drops from 40 miles to 23 miles; and number of trips per person decreases from 3·7 (3·1 non-work trips and 0·6 work trips) to 2·7 (2·3 non-work trips and 0·4 work trips) trips (Figure 2). One can note that the mobility trends change rapidly around March 13 when the national emergency is declared, which is indicated by a grey bar in Figure 2, in accordance with the rapid increase of COVID-19 cases. This observation could have occurred since the emergency declaration raised public awareness on the pandemic, helped the wider spread of the information related to COVID-19, and encouraged more people to reduce mobility.

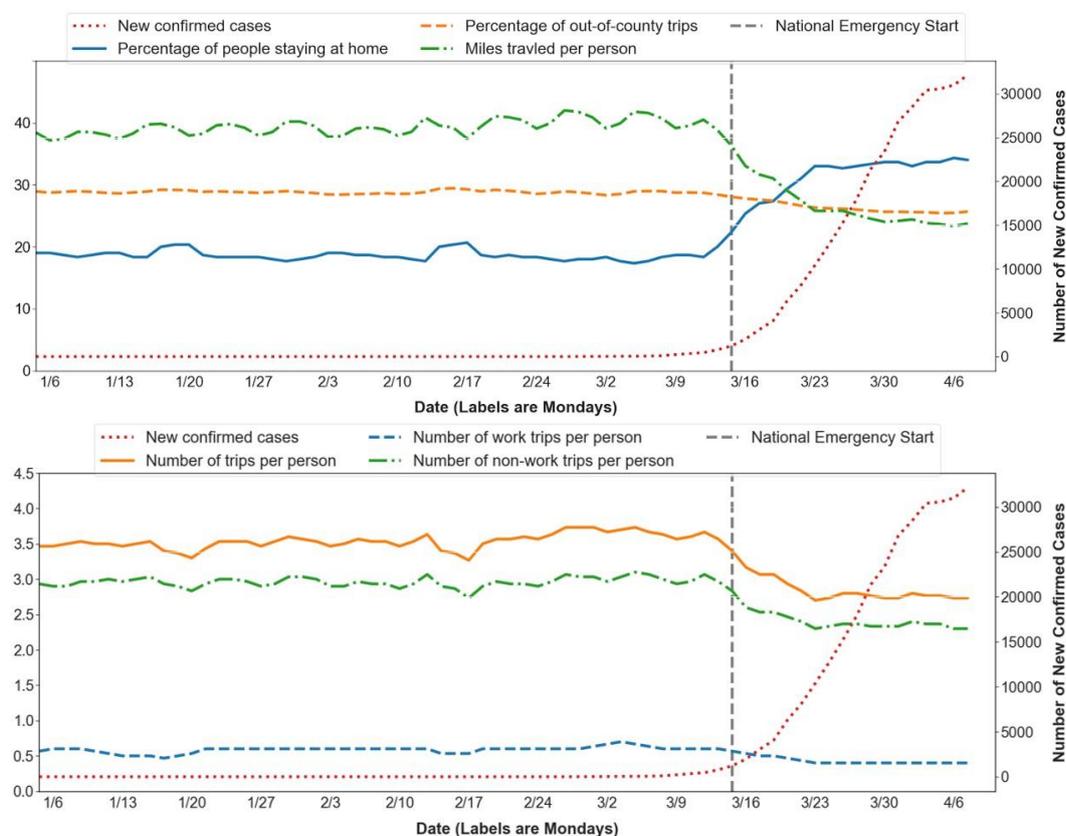

*Figure 2:* **National Trend on Mobility Measurements**

**Statewide trends**
Figure 3 shows the percentage of people staying home in highest order on the x-axis, while states without the mandate are deemphasized by the grey shade. In the most recent week, the District of Columbia maintains the highest rate of people staying home (54%), followed by New York (49%), and New Jersey (45%). Three states with the lowest rates are Mississippi (27%), South Carolina (27%), and Arkansas (26%). In terms of changes between the week before and after the order, three states with highest increase are New Jersey (+13%), New York (+11%), and Illinois (+11%) as marked with a green box in Figure 3. The lowest changes belong to Kentucky (+1·2%), Maine (+0·7%), and South Carolina (-0·2%). The average percentage increase is +4·3% between one week after and one week before order.

Stay-at-home orders also result in trip distance reduction (Figure 4). Hawaii records the lowest miles traveled per person (4·9 mi.), followed by The District of Columbia (15·8 mi.), and Rhode Island (17·2 mi.). Three states with the highest miles traveled are Wyoming (39·0 mi.), Utah (31·4 mi.), and New Mexico (30·1 mi.). The highest decline after the order is observed in Illinois (-9·7 mi.), California (-9·0 mi.), and New Jersey (-8·9 mi.), which are marked with a green box in Figure 4, while the lowest changes are found in Missouri (+1·9 mi.), South Carolina (+0·9 mi.), and Pennsylvania (-0·3 mi.). The stay-at-home order leads to four miles distance reduction on average.

To further evaluate the public reaction stability, we choose the percentage of people staying home statistics. There emerges a larger temporal variance in the reactions of all the states after mid-March, possibly due to various spread status, while the average of populations staying home has become higher (Figure 5). Another approach applied is comparing the statewide public reaction stability based on two different pandemic stages. The first stage is between the temporal range from the date of the first COVID-19 case confirmed to the stay-at-home mandatory order issued in each state (Figure 6(a)). The second stage is after the order was released (Figure 6(b)). For states without the mandates, the first stage ends with our study period (the most recent week). In general, all states in the first stage show higher variances than in the second stage. It implies that the public had been taking active responses, voluntarily staying home more, to the in-state confirmed cases even before mandatory orders were issued. The stay-at-home mandates are observed to be closely stabilizing the statewide movements.

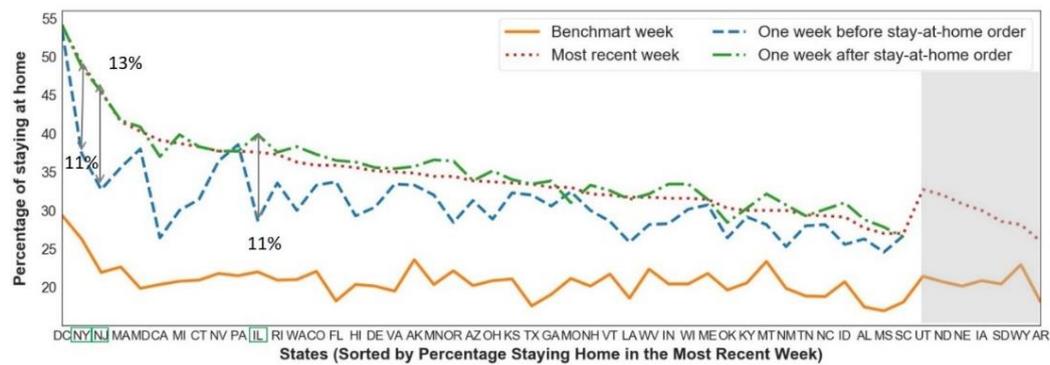

*Figure 3:* **State Trends on Percentage of People Staying Home**

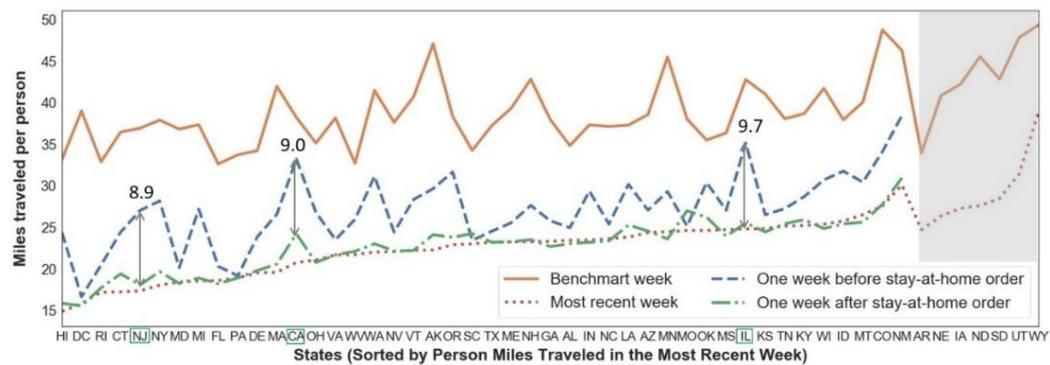

*Figure 4:* **State Trends on Miles Traveled per Person**

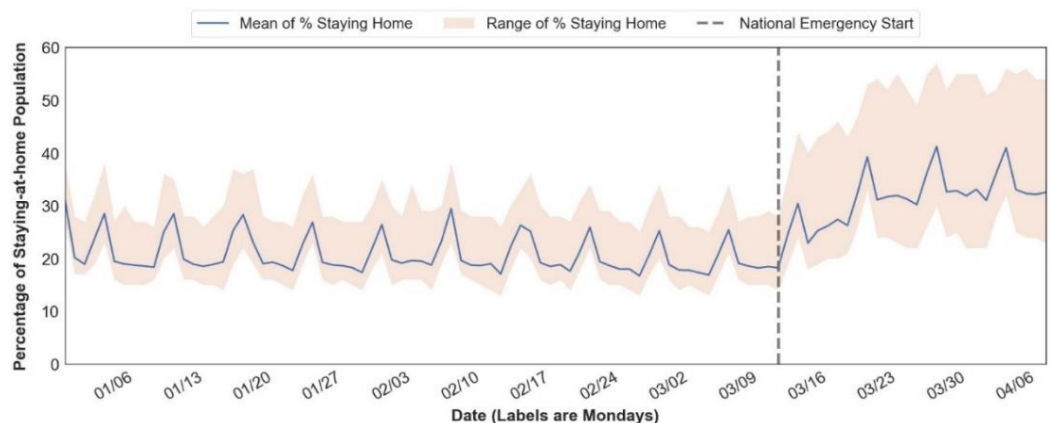

*Figure 5:* **Temporal Variations on Percentage of People Staying Home for All States**

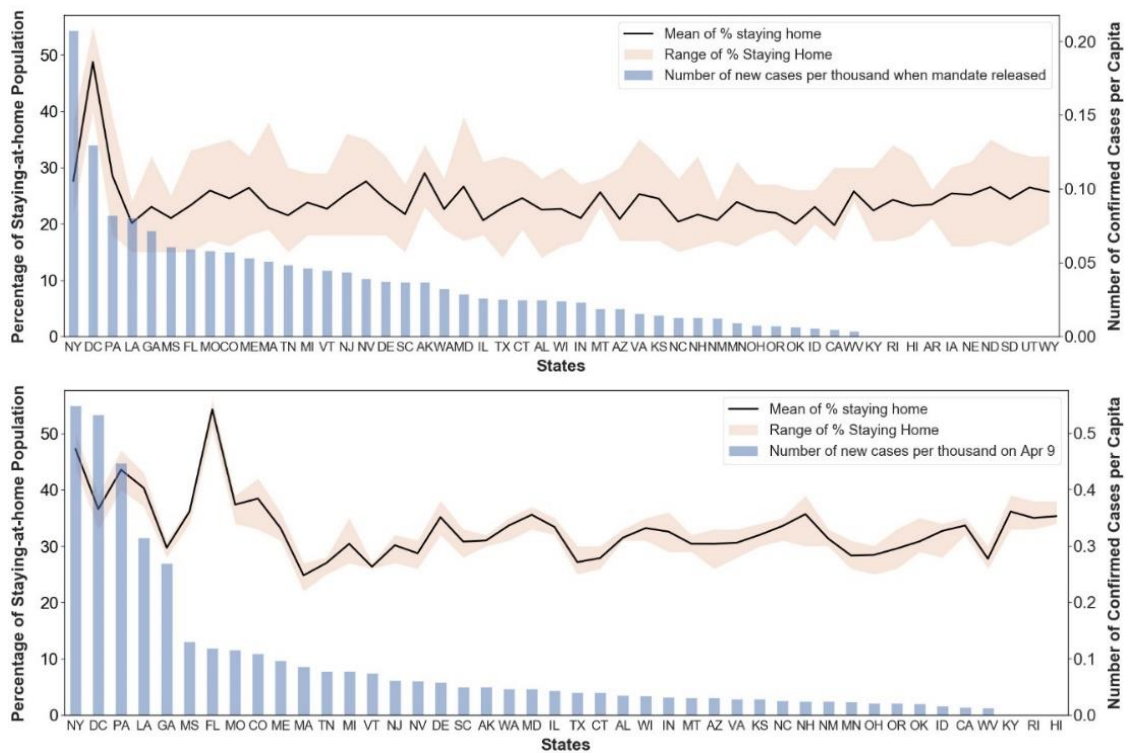

*Figure 6:* **(a) Percentage of People Staying Home between the First Confirmed Case and Stay-at-home Order (*upper*); (b) Percentage of People Staying Home after the Stay-at-home Order (*lower*)**

**Sociodemographic groupwise trends**

Next, we further demonstrate the mobility trends featured by sociodemographic groups. Median income level and population density show a notable clustering nature, which we choose to deliver in this paper. It is of great importance to mention that any features should not be singled out as the only contributing factor without theoretical grounds and rigorous research. Yet, several interesting findings are observed in relation to COVID-19.

Figure 7(a) illustrates the percentage of people staying home for income group comparison (upper), and Figure 7(b) demonstrates miles traveled per person in the order of population density (lower). The mobility pattern gradually becomes distinguishable between the two groups upon the COVID-19 outbreak. Until the national emergency declared on March 13 (grey line), states share a relatively homogeneous staying home trend regardless of the sociodemographic features, while one can pinpoint a more distinctive trend thereafter. The higher income group tends to present a higher staying-at-home ratio and higher density group present lower trip distance after the outbreak. Even though the income level or population density may not be the only factors that affect the mobility trend, there exists overall heterogeneity between the groups.

Figure 8 illustrates groupwise mobility patterns in three measures. One point to note is that two income groups exchange their trip distance trend around mid-March (grey bar) (Figure 8(a)). The high-income states tend to travel longer distances before the pandemic, while the low-income group surpasses later. The higher income group tends to stay at home slightly more before the outbreak, but the gap becomes more notable afterwards. The percentage of out-of-county trips relatively stays similar between groups. However, the high-income group traveled slightly more to out-of-county locations, whereas the difference becomes obscure after mid-March. The percentage of people staying home trend also exhibits a similar pattern until mid-March (grey bar) between two population density groups (Figure 8(b)). Afterwards, the higher density group reduces their mobility noticeably. One might assume that the higher chances of contacting other people in the higher density area result in practicing social distancing more actively. Both before and after the outbreak, the lower density states sustain higher miles traveled and lower out-of-country trip rates. Different from income comparison, no metrics exchange the trends and, instead, the gaps in miles traveled and out-of-county trips are more noticeable between the groups throughout the whole study period.

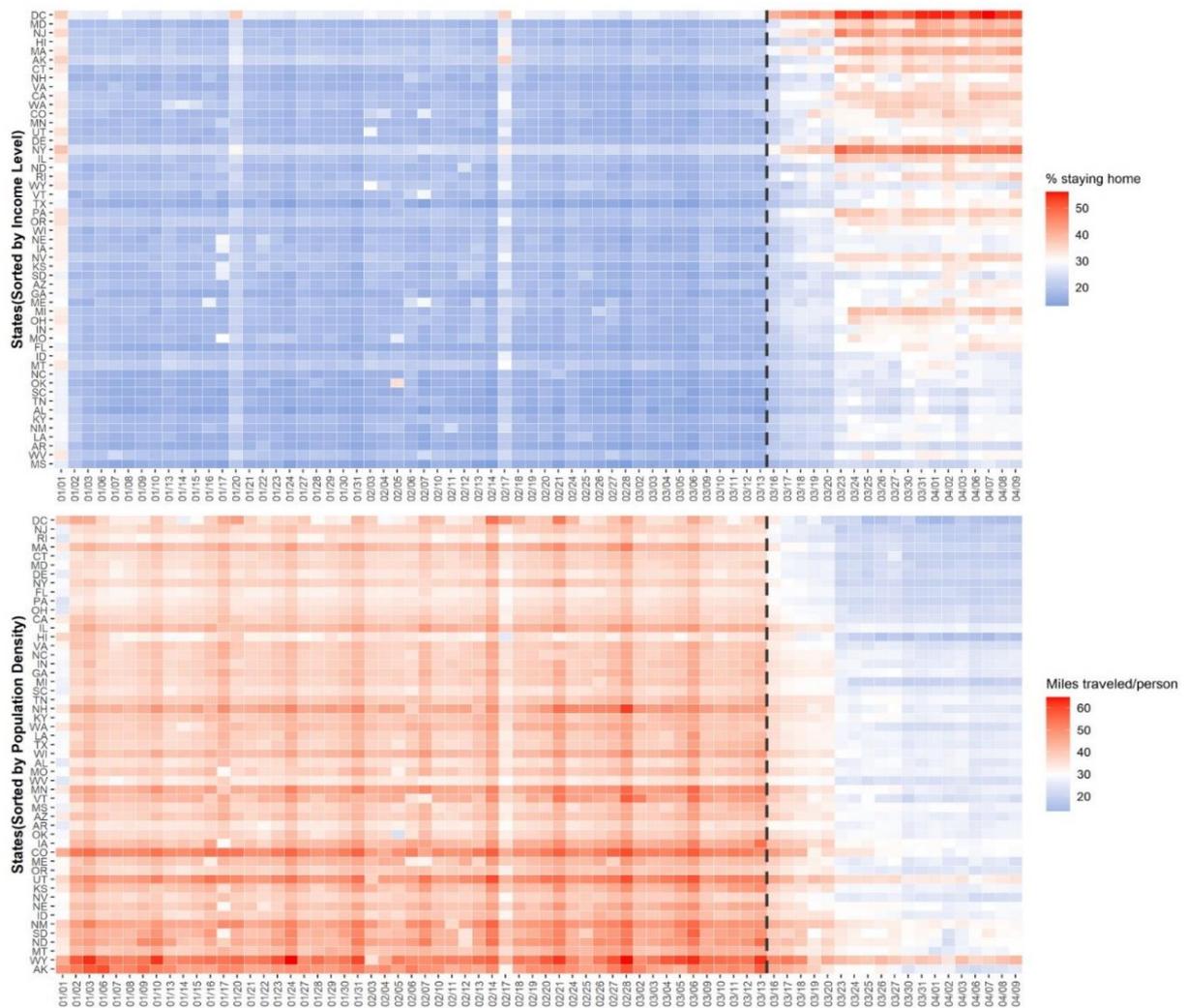

*Figure 7:* **(a) Spatiotemporal Difference in Percentage of People Staying Home by Income Order (*upper*); (b) Spatiotemporal Difference in Miles Traveled per Person by Population Density Order (*lower*);**

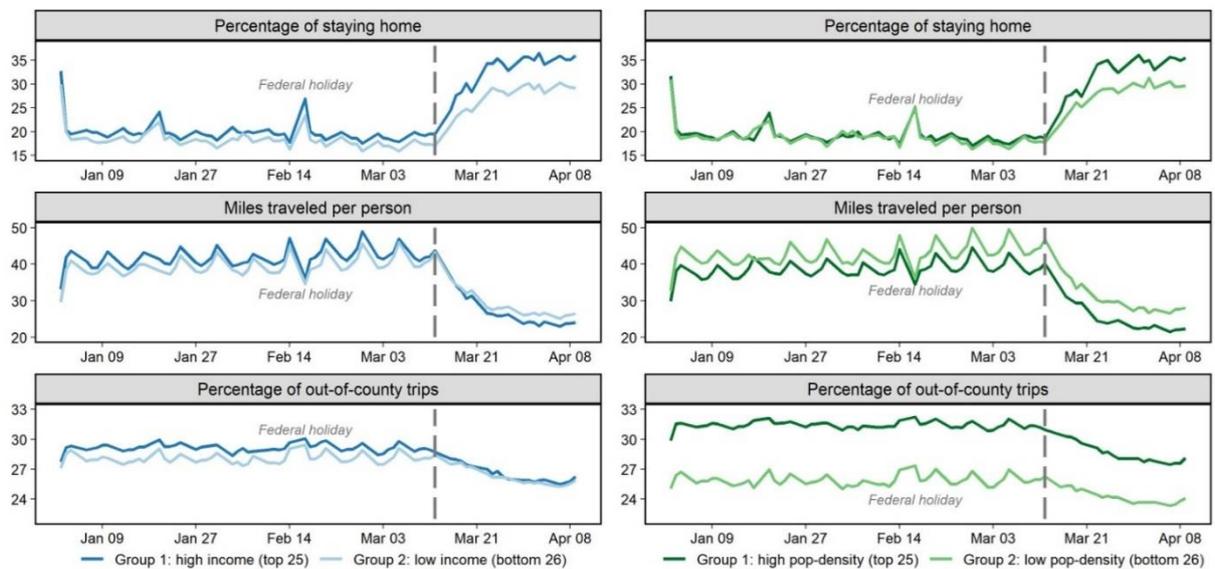

*Figure 8:* **(a) Mobility Patterns by Income Groups (*left*); (b) Population Density Groups (*right*)**

## Discussion

Understanding public reactions to the virus and the non-pharmaceutical interventions should be of great help to fight COVID-19 in a strategic way. In order to provide an integrated perspective on public reactions related to the pandemic propagation and the non-pharmaceutical interventions, we examine the day-by-day mobility variations across the U.S. by leveraging mobile device location data and the measures related to social distancing. While data-driven study we conduct shows the mobility patterns for the general public, we would like to introduce an additional potential research area that utilizes the mobility measures as an open discussion.

**Unemployment Issue during the Pandemic**

A teleworking rate is another important measure to reveal how many employees have been impacted from COVID-19. As discussed, the work trip rates decrease amid the pandemic (-0·2 work trips per person) resulting from both teleworking and unemployment increase. Here we attempt to understand the increased rate of people staying home whether it is due to teleworking or unemployment by estimating the teleworking rate. This estimate is based on the work trip frequency per person and weekly unemployment claims from the United States Department of Labor.[33] For this demonstration, we first divide employees into two categories: teleworkers and commuters. Their initial relative ratio is estimated per state based on the American Community Survey (ACS) report.[34] Total number of employees ($teleworkers + commuters$) in the benchmark week are obtained based on the weekly unemployment claims. Commuters in the benchmark week are estimated as the product of commuter ratio and total employees. Assuming that the work trip rate per person is consistent within a short time period, we calculate the number of commuters by dividing the number of work trips by the work trip rate per person. Then, the number of teleworkers per week is estimated ($total\ employees - commuters - unemployed\ workers$) with an assumption that there is no additional weekly employment during COVID-19. Here, the number of weekly unemployed populations are estimated based on the unemployment claims.

Despite our naive approach and assumptions, the estimates still conform with the pandemic circumstances. The teleworking rate starts to increase around mid-March complying with the sharp increase in COVID-19 cases, in which the baseline is set on the week of March 7 (Figure 9). As of April 4, the number of employees working from home increases by 25%, compared to the baseline week. The teleworking rate in some states stops increasing and even decreases in April. Two possible assumptions account for this trend: employees who no longer conduct teleworking and work on site or are no longer hired. One drawback to our estimation is that it does not separate the population who are on a break from teleworkers, which is the possible reason for the higher teleworking rate in the first week of January. Overall, we anticipate more research is invited to scrutinize this teleworking trend regarding COVID-19 and rapidly inclining unemployment rates, which marks the highest increase in the seasonally adjusted series report history.[35] In addition, as the teleworking trend has not been observed to this extreme extent before, further research will help provide essential evidence on the economic crisis.

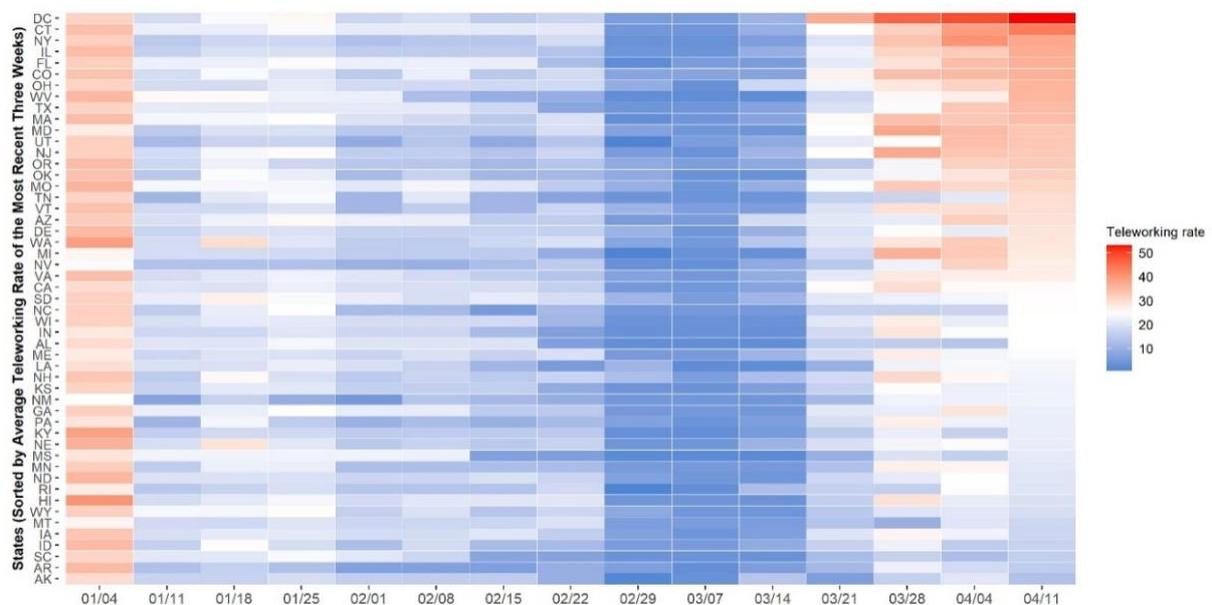

*Figure 9:* **Teleworking Trends for Employees**

**Findings**

Based upon the noticeable degree of mobility trend alteration at a nation level, our study provides more detailed evidence on the mobility shifts from the statewide and groupwise comparison. Evidenced by our measures, a large number of populations have decreased their daily movements during the pandemic. This may suggest that the government message urging to stay home conforms with the public mobility behavior. Our findings can be summarized as follows, which are closely related to social distancing trends:

- the nationwide mobility has reduced amid the pandemic observed by all metrics;
- the declaration of the national emergency with the rapid increase in COVID-19 cases can be perceived as a significant stimulus to the increase in people staying home and decrease in mobility;
- the public has been taking active responses, voluntarily staying home more, to the in-state confirmed cases while the stay-at-home mandates stabilize the staying-at-home population with a smaller range of fluctuation;
- even though the income level or population density may not be the only factors that affect the mobility trend, there exists overall heterogeneity between the groups;
- income groups exchange the trend of miles-traveled-per-person measure, such that the higher income group tends to travel longer distance before the pandemic while the low-income group surpasses the trend afterwards;
- both before and after the outbreak, the lower density states sustain higher miles traveled and lower out-of-country trip rates and no metrics exchange the trends between density groups; and
- the gaps in miles traveled and out-of-county trips are more noticeable between the population density groups throughout the whole study period, compared to the income groups.

One limitation in our study is that we are not certain at which pandemic stages of the public mobility reactions are demonstrated. This part can be explained when the pandemic vanishes, hopefully in near future. Now with many more weeks into the pandemic from the study period, there could be an opportunity to find relationships between the plateau of confirmed cases and that of social distancing trends. In addition, as we observe the clear evidence on social distancing efforts, a rigorous modeling approach will be necessary to quantify the social distancing trends and to analyze the reasons behind. Also, our observation could be also integrated with the pharmaceutical modeling research. While carefully suggesting the potential and necessary research areas that can greatly help the current and potential future matters, we anticipate our data-driven analysis offers integrated insights on human mobility trends regarding the pandemic circumstances. This study could serve as evidence to raise public awareness and, consequently, reinforce the importance of social distancing while assisting policymakers.

## Acknowledgments

We would like to thank and acknowledge our partners and data sources in this effort: (1) Amazon Web Service and its Senior Solutions Architect, Jianjun Xu, for providing cloud computing and technical support; (2) computational algorithms developed and validated in a previous USDOT Federal Highway Administration's Exploratory Advanced Research Program project; (3) partial financial support from the U.S. Department of Transportation's Bureau of Transportation Statistics; and (4) mobile device location data from several data provider partners.